\documentclass[acmsmall,screen,nonacm]{acmart} 
\AtBeginDocument{%
  }

\setcopyright{acmlicensed}
\copyrightyear{2018}
\acmYear{2018}
\acmDOI{XXXXXXX.XXXXXXX}

\acmJournal{POMACS}
\acmVolume{37}
\acmNumber{4}
\acmArticle{111}
\acmMonth{8}
\usepackage{tikz}
\usetikzlibrary{arrows.meta}
\usepackage{booktabs}
\usepackage{longtable}
\usepackage{array}
\usepackage{pgfplots}
\pgfplotsset{compat=1.18}
\usetikzlibrary{positioning, arrows.meta, shapes.geometric, fit}
\usepackage{tabularx}
\usepackage{placeins}
\definecolor{TrustBlue}{HTML}{0072B2}
\definecolor{AutonomyOrange}{HTML}{D55E00}
\definecolor{PrivacyGreen}{HTML}{009E73}
\definecolor{SatisfactionPurple}{HTML}{CC79A7}

\author{Liqian You}
\email{Liqian.You@student.uts.edu.au}
\authornotemark[1]
\affiliation{%
  \institution{University of Technology, Sydney}
  \city{Sydney}
  \state{NSW}
  \country{AU}
}

\author{Jianlong Zhou}
\affiliation{%
  \institution{University of Technology, Sydney}
  \city{Sydney}
  \country{AU}}
\email{Jianlong.Zhou@uts.edu.au}

\renewcommand{\shortauthors}{Liqian et al.}

\begin{document}

\title{Designing Automation Boundaries for Trustworthy Smart Medication Support}


\renewcommand{\shortauthors}{}

\begin{abstract}
Smart medication systems increasingly automate medication recognition, reminders, and logging. However, automation in home medication routines should be carefully bounded, as users may have different capabilities, privacy expectations, and needs for control over decisions. We present a mixed-methods study of a Smart Medication Support system comparing three automation conditions: confirmation required, automatic logging with undo, and fully automatic support. Across 53 participants and interviews with 11 older adults, we found that higher automation did not necessarily lead to higher trust or acceptance. Participants preferred automation that reduced routine effort while preserving opportunities for correction. Fully automatic support was less interruptive but was rated lower in autonomy, trust, transparency, dignity, and satisfaction. Interviews also showed clear differences among older adults. Their preferences were shaped by privacy concerns, digital confidence, perceived vulnerability, and caregiver involvement. We contribute empirical evidence and design implications for calibrating automation in smart medication systems according to task risk, user control, and ethical acceptability.\end{abstract}

\ccsdesc{Ubiquitous and mobile computing}
\ccsdesc{Life and medical sciences}
\ccsdesc{Human-computer interaction (HCI)}

\keywords{Smart home, ethical acceptance, trust}

\received{20 February 2007}
\received[revised]{12 March 2009}
\received[accepted]{5 June 2009}

\maketitle

\section{Introduction}

Smart home health technologies are increasingly positioned as a way to support aging in place \cite{liu2016smart}, chronic disease management \cite{vranvcic2024role}, and everyday care in domestic settings \cite{tian2024benefits}. Among these applications, medication management \cite{varshney2013smart} is particularly important. For older adults and people living with chronic conditions, taking the right medicine at the right time is a recurrent and consequential care activity. This activity is often complicated by polypharmacy, changing prescriptions, memory demands, and the need to coordinate with caregivers or clinicians \cite{majumder2017smart}. Smart medication systems have moved beyond simple reminders to include medicine recognition, automated logging, remote monitoring, and safety alerts \cite{faisal2023key,arain2021medication}. As these systems become more capable, the key design challenge shifts. The question is no longer whether medication support can be automated, but how automation should be calibrated for safe, trustworthy, and acceptable use in everyday care.

Calibrating automation \cite{pop2015individual} is important because medication support is not only about efficiency. When a system recognizes a pill, records intake, escalates reminders, or interrupts a user who selects the wrong medicine, it redistributes initiative between the user and the system. Prior work on automation and human factors has shown that automation shapes not only performance, but also reliance, oversight, and recovery from error \cite{lee2004trust,sarter1997team}. In domestic healthcare, these issues become especially sensitive because the home is a personal space and medication is closely tied to safety, independence, and routine. An automatic intervention may reassure users when it helps prevent harm, but it may also feel intrusive when it interrupts familiar routines, records actions without clear consent, or leaves users uncertain about when to rely on the system and when to intervene.

Trust is a central concern in the design of smart medication systems \cite{you2025behavior}. In this context, trust involves more than confidence in technical reliability. It also depends on whether users understand what the system is doing, whether warnings are timely and appropriate, and whether users retain meaningful opportunities to confirm, correct, or override system actions \cite{lee2004trust}. Lower automation may preserve user control, but it can also increase workload and reduce the perceived value of the system \cite{dawadi2013automated}. Higher automation may provide convenience and protection, but it can also reduce transparency and perceived agency when actions are executed without confirmation \cite{sarter1997team}. These tensions suggest that trust does not depend only on the amount of automation. It also depends on the medication task, the perceived risk of error, and the forms of control available to users.

Existing literature \cite{pirzada2022ethics,felber2023mapping,ji2022scoping} provides only a partial account of these tensions. Research on smart home healthcare for older adults has identified privacy, autonomy, dignity, trust, usability, and fairness as important determinants of acceptance \cite{zhu2022ethical,tian2024benefits,ji2022scoping}. Studies of older adults' technology adoption further show the importance of systems being understandable \cite{shuhaiber2019understanding}, acceptable within everyday routines \cite{ghorayeb2021older}, and respectful of users' values \cite{dermody2021factors}. However, much of the medication literature focuses on reminders, adherence support, product features, or technical feasibility, rather than examining automation level as a design variable in its own right \cite{faisal2023key,patel2020prospective,ahmad2020users,arain2021medication}. Conversely, broader work on automation and ethics highlights appropriate reliance \cite{lee2004trust}, meaningful human involvement, and control \cite{santoni2018meaningful}, but rarely examines how these concerns unfold in concrete medication tasks at home \cite{davidovic2023purpose}. We lack understanding of how different allocations of initiative between user and system shape trust in smart medication support, and how this trust relates to concerns about privacy, autonomy, dignity, and comfort.

To address this gap, we investigate automation boundaries in a Smart Medication Support system through a mixed-methods HCI study. We examine automation boundaries as a design choice about when the system acts, when it asks, and when users remain in control. We compare three levels of automation: user confirmation before action, automatic action with an undo option, and fully automatic routine support. These levels across three medication scenarios vary in routine and risk: on-time intake, missed or delayed medication, and wrong medicine selection. To capture both behaviour and experience, we combine questionnaire ratings, preference rankings, interviews, and observation notes, with attention to participants' responses to reminders and system warnings. This allows us to examine how automation levels shape trust and related experiential and ethical concerns, including privacy, autonomy, dignity, and comfort.

This paper contributes:

\begin{itemize}
    \item A conceptual model of automation boundaries in domestic healthcare, linking automation level, task context, user control, and perceived ethical acceptability.
    \item Empirical evidence showing that trust does not simply increase with automation, but varies across medication scenarios and forms of user control.
    \item A design framework for calibrating automation in smart medication systems, outlining when systems should automate, request confirmation, or intervene.
\end{itemize}

\section{Related Work}

\subsection{Automation Boundaries in Smart Medication Management}

Smart home health technologies increasingly support aging in place, chronic disease management, and domestic care \cite{liu2016smart,majumder2017smart,vranvcic2024role}. Medication management is a frequent and consequential task in these settings \cite{varshney2013smart}. Early work mainly framed medication support as reminders, scheduling, and adherence assistance \cite{varshney2013smart}. Recent systems extend this model through intelligent dispensers \cite{faisal2023key}, automated logging \cite{turjamaa2020smart}, caregiver notifications \cite{peddisetti2024smart}, medicine verification \cite{verdu2025smart}, and locking mechanisms \cite{kapse2025mediserve}. IoT systems now combine dispensers, smart cups, mobile alerts, and consumption checks \cite{peddisetti2024smart}. Devices such as MyAide Smart Dock record medication removal, timing, and user identity \cite{wallace2025validation}. HCI work has also explored AIoT support for older adults, including HUG Smart Sticker for personalised medication management \cite{hao2025hug}. These studies show a shift from reminder tools to systems that sense, record, notify, and act within medication routines.

Most existing studies still focus on adherence, usability, workload, accuracy, or feasibility \cite{ahmad2020users,arain2021medication,patel2020prospective,faisal2023key}. Work on real-time adherence monitoring shows that medication data can support patients, caregivers, clinicians, pharmacists, and payers, while also raising issues of data interpretation, monitoring responsibility, and everyday use \cite{faisal2023exploring}. Less work treats automation level itself as a design variable: when the system should act, when it should ask, and when users should retain control. This matters because medication tasks differ in routineness, urgency, reversibility, and consequence severity. The same automation may be helpful in routine use, intrusive in personal decisions, and unsafe in high-risk cases. We therefore examine automation boundaries: how system initiative, confirmation, and correction shape user experience and ethical acceptance.

\subsection{Trust Calibration and Appropriate Reliance}

Automation research shows that more trust is not always better. Lee and See \cite{lee2004trust} argue that trust should support appropriate reliance: users should rely on automation when it is capable, but remain able to monitor, question, or reject it. Too little trust can cause under-reliance. Too much trust can cause over-reliance. Work on complex automated systems also shows that higher capability can create coordination problems, including automation surprises and reduced oversight \cite{sarter1997team}.

These concerns are important for medication support \cite{tolley2018improving}. Smart medication systems generate records through sensors, apps, dispensing logs, and caregiver monitoring \cite{peddisetti2024smart,wallace2025validation,faisal2023exploring}. These records can support timely care, but they may also be read as stronger evidence than they are. For example, medication access may be mistaken for medication ingestion. Confirmation can preserve oversight, but it may add effort in routine use. Fully automatic support can reduce effort, but it may leave users out of the decision loop and make errors harder to notice or contest. Auto with undo offers a middle position: routine actions proceed with less effort, while users keep a recovery point after system action. Undo is more than error recovery. In medication support, it may also help calibrate reliance by making automated actions less final and more recoverable \cite{wischnewski2023measuring,de2020towards}. XAI work similarly argues that trust evaluation should distinguish between trusting correct outputs and over-trusting incorrect ones \cite{kok2023explainable,das2023explainable,miro2026toward}. Our study does not test algorithmic correctness directly. Instead, it examines how confirmation, undo, and full automation affect users' perceived ability to rely on, monitor, and correct system actions.

\subsection{Meaningful Human Control and Task Risk}

Work on meaningful human control argues that human involvement matters for safety, accountability, dignity, and contestability \cite{santoni2018meaningful,davidovic2023purpose}. In medication support, these concerns are practical. A system may record intake, warn about a wrong medicine, change a schedule, or share information with a caregiver. Each action shifts some responsibility between the user and the system.

Meaningful control requires people to track and intervene in system behaviour \cite{santoni2018meaningful}. For medication automation \cite{peddisetti2024smart}, this suggests that control mechanisms should vary by task. Routine and low-risk actions may need clear feedback. Routine but consequential actions may need undo or review. Rare or high-risk actions may need explicit confirmation before execution.

We use task routineness and consequence severity to describe these boundaries. Table~\ref{tab:task_risk_matrix} summarises this logic. Automation should not depend on system capability alone. It should also depend on whether users can understand, contest, and recover from system action. Simple reminders may be automated. Intake logging may require undo. Wrong-medication warnings, dosage changes, prescription updates, and data sharing should require stronger confirmation because they are higher risk or more tied to personal responsibility. Verbeek's work on ambient intelligence further shows that automated systems can shape behaviour even when users do not fully notice them \cite{verbeek2009ambient}. In medication support, system correctness and user awareness are both central design concerns.

\begin{table}[ht!]
\centering
\small
\caption{Task-risk logic for setting automation boundaries in smart medication support.}
\label{tab:task_risk_matrix}
\begin{tabular}{p{2.5cm}p{5cm}p{5.4cm}}
\toprule
& \textbf{Low consequence severity} & \textbf{High consequence severity} \\
\midrule
\textbf{High routineness}
& Automate with clear feedback
& Automate with immediate undo or review \\
\textbf{Low routineness}
& Automate with review or explanation
& Require explicit confirmation before action \\
\bottomrule
\end{tabular}
\end{table}

\subsection{Ethical Acceptance in Smart Home Care for Older Adults}

Appropriate reliance and task risk explain how automation can be calibrated for effective use. In domestic healthcare, however, effectiveness is not enough. Users also need to see automation as proper, respectful, and acceptable in everyday care. This broader concern is captured by ethical acceptance \cite{pirzada2022ethics,zhu2022ethical,felber2023mapping}.

Ethical acceptance is central in smart home care because these systems operate in private spaces, use health data, and often support older adults with changing physical, cognitive, or social needs. Prior work identifies privacy \cite{zheng2018user}, autonomy \cite{klugman2018ethics}, dignity \cite{joseph2025designing}, social support \cite{lee2017companionship}, and perceived surveillance \cite{percy2024user} as key concerns in older adults' technology acceptance \cite{felber2023mapping,ji2022scoping}. Ethics-by-design work further shows that these values are shaped by concrete choices, including sensing, interface design, alert thresholds, data access rules, and options for override \cite{tonkinwise2004ethics,hine2022ethical,wang2023proactive}.

Recent HCI and digital health work makes this point in medication and health routines. HUG Smart Sticker designs AIoT medication support around older adults' routines and needs \cite{hao2025hug}. Work on voice assistants highlights personalisation, context adaptation, and respect for autonomy in health self-management \cite{mahmood2025voice}. Studies of real-time adherence monitoring show that medication data can support communication among patients, caregivers, clinicians, and pharmacists, while raising questions about access, interpretation, and management \cite{faisal2023exploring}.

In medication support \cite{tolley2018improving}, these values appear in small interaction choices: whether the system asks before recording intake, whether users can undo an automatic log, whether data sharing requires consent, and whether warnings are framed as support rather than judgement. Older adults also differ in digital confidence, privacy expectations, perceived need, and desire for customisation \cite{ghorayeb2021older,dermody2021factors}. Automation boundaries cannot be set only at the system level. What feels supportive for one user may feel intrusive or insufficiently protective for another.

Smart home care technologies may reduce caregiver burden and support independent living, but they may also make users feel watched or judged \cite{birchley2017smart,tian2024benefits}. In medication support, the ethical question is not only how much automation users prefer. It is how automation changes control, responsibility, and visibility among older adults, caregivers, clinicians, and the system. Our study addresses this gap by comparing confirmation required, auto with undo, and fully automatic support within the same medication context.

\section{Prototype Design: Smart Medication Support}

\subsection{System Overview}

The Smart Medication Support system was developed as an interaction probe for studying automation boundaries in home medication support. It was not designed as a technical contribution to recognition performance or hardware engineering. Instead, it provided a controlled platform for presenting different levels of system initiative in comparable medication tasks. These conditions corresponded to confirmation required, auto with undo, and fully automatic support.

The system consisted of two integrated components. The mobile application managed medication records, schedules, and user guidance. The situated interaction interface supported medication recognition, reminder responses, confirmation, undo, and feedback presentation when required by the active automation condition. Together, these components created a realistic medication support context in which user control and system agency could be varied systematically.

\begin{figure}[ht]
\centering
\includegraphics[width=\columnwidth]{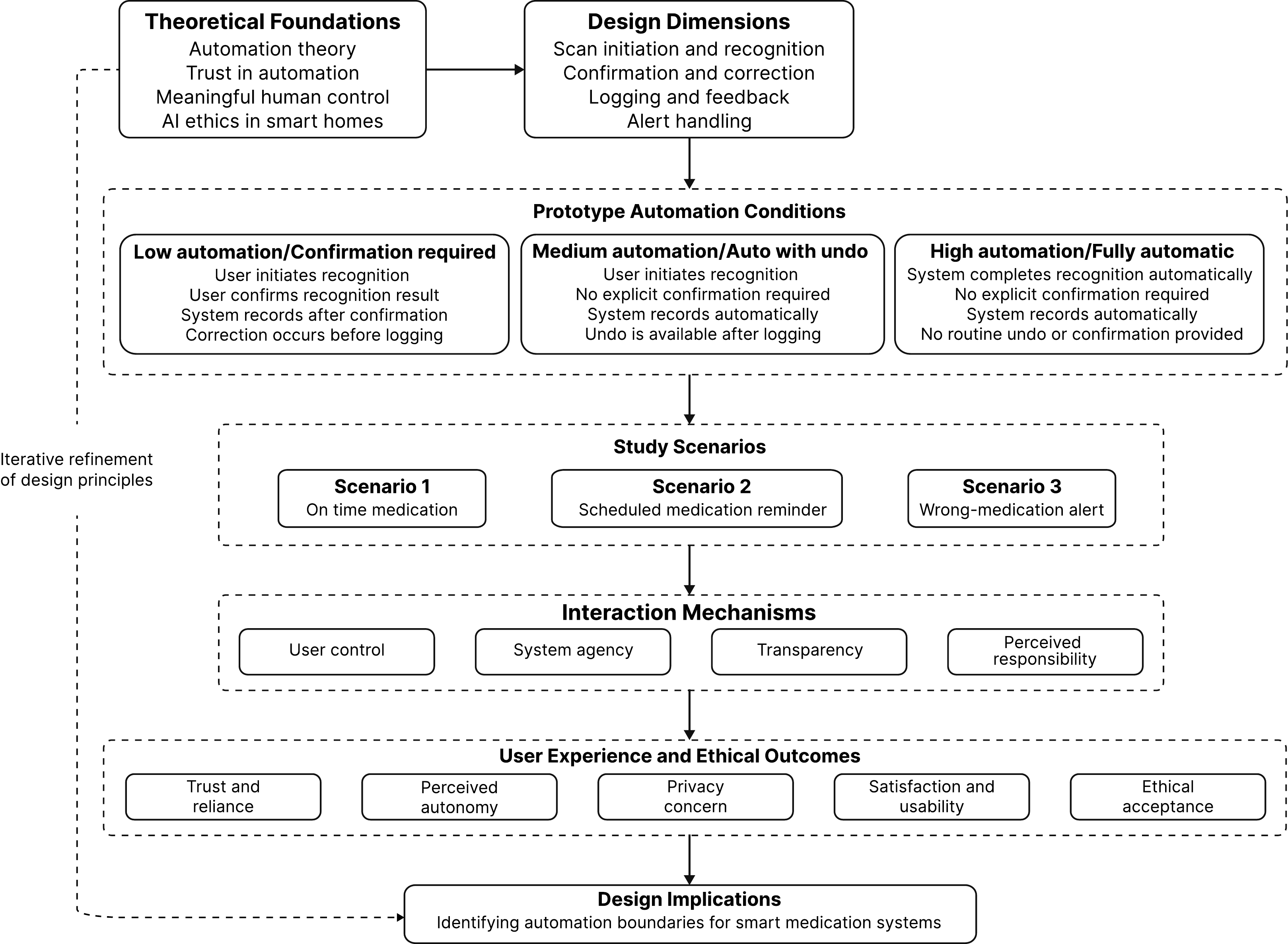}
\caption{Conceptual design framework for examining automation boundaries in smart medication systems.}
\label{fig:design_framework}
\end{figure}

As shown in Figure~\ref{fig:design_framework}, the study conceptualises automation level as a design variable. It shapes user control, system agency, transparency, and perceived responsibility. These mechanisms are examined in relation to trust and reliance, autonomy and control, privacy concern, satisfaction, usability, and ethical acceptance.

The study held the recognition process and feedback presentation consistent across automation conditions. The main variation was how control was allocated between the user and the system. This included who initiated recognition, whether confirmation was required, whether intake logging occurred automatically, and whether correction was available after system action.

The system supported three medication scenarios: on time medication, missed or delayed medication, and wrong medication detection. Each scenario was presented under the three automation conditions. This design allowed the study to compare participant responses across routine medication tasks and safety related situations.

\subsection{Prototype Architecture}

We developed the Smart Medication Support system as an interaction probe for studying automation boundaries in home medication support. The system was not designed to advance recognition accuracy or infrastructure. It was designed to support a controlled comparison of three automation conditions while keeping the core recognition process consistent across tasks.

\begin{figure}[ht]
\centering
\includegraphics[width=\columnwidth]{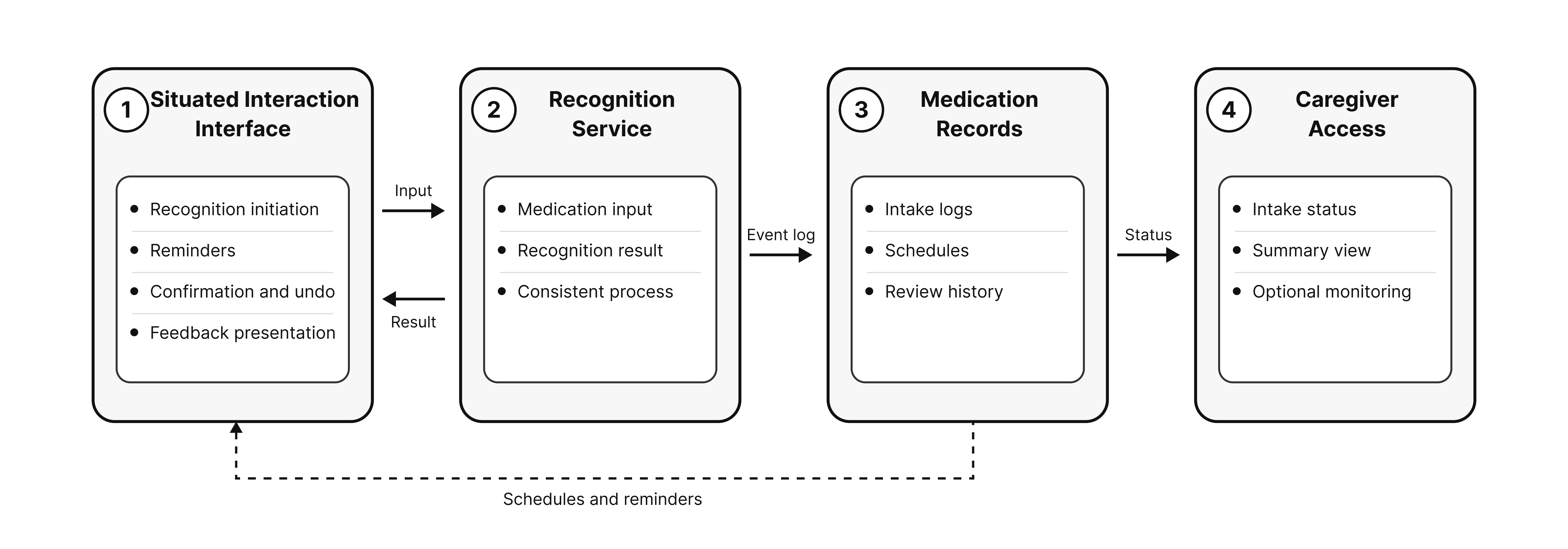}
\caption{Functional architecture of the Smart Medication Support system.}
\label{fig:design}
\end{figure}

Figure~\ref{fig:design} shows four functional components of the system: the situated interaction interface, recognition service, medication records, and caregiver access. The situated interaction interface supported recognition initiation, reminders, feedback, confirmation, and undo when required by the active condition. The recognition service processed medication inputs and returned recognition results for display and logging. Medication records stored intake logs and schedule information. Caregiver access provided a read-only view of intake status and summary for monitoring.

The same recognition process and feedback structure were used in all three conditions. The study therefore did not compare different levels of accuracy, system performance, or interface layout. The experimental variation was how control was allocated between the user and the system. This included who initiated recognition, whether confirmation was required, whether intake was logged automatically, and whether users could undo or review an action after logging.

This architecture allowed the study to isolate automation boundary as the main design variable. Low automation preserved explicit confirmation before logging. Medium automation reduced routine effort while keeping an undo option. High automation reduced routine user input during recognition and logging. By keeping the recognition process stable, the prototype supported a controlled comparison of trust, autonomy, privacy concern, comfort, and ethical acceptance across automation conditions.

\subsection{Automation Conditions}

To examine how different degrees of automation shape user experience and ethical acceptance, we operationalised automation level as the allocation of initiative, confirmation, logging, and recovery between the user and the system. The Smart Medication Support system was implemented in three ordered automation levels: Condition A, low automation / confirmation required; Condition B, medium automation / auto with undo; and Condition C, high automation / fully automatic. These labels refer to the degree of system agency in routine medication recognition and logging, rather than to differences in recognition accuracy, hardware performance, or interface layout. Across conditions, the interface structure, recognition process, reminder functions, and feedback channels were kept consistent. The experimental manipulation was whether users initiated recognition, whether they confirmed results before logging, and whether they could recover from system action through undo or review.

\begin{table}[ht!]
\centering
\footnotesize
\caption{Automation levels and interaction conditions implemented in the Smart Medication Support system.}
\label{tab}
\begin{tabular}{p{0.5cm}p{1.5cm}p{1.5cm}p{2cm}p{2.5cm}p{3.8cm}}
\toprule
\textbf{Label} 
& \textbf{Automation level} 
& \textbf{Interaction condition} 
& \textbf{Recognition initiation} 
& \textbf{Confirmation} 
& \textbf{Logging and recovery} \\
\midrule
A 
& Low automation 
& Confirmation required 
& User initiates recognition. 
& User reviews and confirms the recognition result before recording. 
& System records the medication event only after confirmation; correction occurs before logging. \\
\addlinespace[2pt]

B 
& Medium automation 
& Auto with undo 
& User initiates recognition. 
& No explicit confirmation is required before logging. 
& System records the medication event automatically; users can undo the logged event after system action. \\
\addlinespace[2pt]

C 
& High automation 
& Fully automatic 
& System initiates and completes routine recognition. 
& No explicit confirmation is requested before logging. 
& System records the medication event automatically; no routine undo or confirmation is provided. \\
\bottomrule
\end{tabular}
\end{table}

In the \textit{confirmation required} condition, the system recognised the medication but did not log the medication event automatically. After recognition, users reviewed the recognition result and confirmed it before the event was recorded. This condition preserved the highest level of user control. It also made system actions visible before they affected the medication record.

In the \textit{auto with undo} condition, recognition and intake logging were completed automatically after the user initiated recognition. The system recorded the medication event without asking for explicit confirmation. It then showed an undo option so that users could reverse the action if needed. This condition reduced routine effort while preserving a recovery point after system action.

In the \textit{fully automatic} condition, the system handled recognition and intake logging with minimal user intervention. Once the medication interaction was initiated, the system completed recognition and logging and provided feedback through the interface. No explicit confirmation was requested, and no routine undo option was provided. Safety-related events, such as wrong medication detection, still produced visible system feedback. This condition represented the highest level of system agency and the lowest level of required user action in routine medication tasks.

These three conditions varied operational control while keeping the broader interaction context stable. This design allowed us to compare how participants responded to different levels of system initiative in terms of trust, perceived control, transparency, and usability.

\section{Method}

We conducted a mixed-methods user study to examine how automation level shaped user experience and ethical acceptance in home medication support. Participants used the Smart Medication Support system under three automation conditions: confirmation required, auto with undo, and fully automatic support. They completed three medication tasks that covered routine use and safety-relevant situations. Quantitative analysis focused on questionnaire ratings, preference rankings, and final comparison items. Qualitative analysis used open responses, structured observation notes, and interviews with 11 older adults. These data helped explain how participants interpreted automation boundaries, trust, privacy, and control.

\subsection{Study Design}

The study used a repeated measures comparative design. Each participant used the same system under three automation conditions: Condition A (low automation / confirmation required), Condition B (medium automation / auto with undo), and Condition C (high automation / fully automatic). The conditions shared the same interface structure, reminder functions, recognition process, and feedback channels. They differed in how confirmation, logging, and recovery were allocated between the user and the system.

Each participant completed the same three medication scenarios within each automation condition: on-time medication, delayed or missed medication, and wrong-medication detection. Automation condition was treated as the primary within-subjects factor. The order of automation condition blocks was counterbalanced across participants to reduce order, learning, and fatigue effects. The scenario sequence was kept fixed within each block as a scripted task progression, moving from routine intake to reminder response and then wrong-medication detection. This allowed each automation condition to be evaluated under the same scenario progression.

During each condition, the researcher recorded structured observation notes. These notes covered task completion, visible hesitation, confusion, reminder responses, unexpected interaction, and observed difficulties with system feedback. After each condition, participants rated usability, trust, autonomy, privacy, comfort, and perceived control. Open responses and short follow-up prompts captured how participants interpreted system behaviour and which forms of automation they considered appropriate for different medication tasks.

A total of 53 participants were recruited through email invitations and community outreach channels. Recruitment targeted adults with experience or interest in medication management, informal care, healthcare, or digital health technologies. We paid particular attention to older adults and caregivers.

\subsection{Tasks and Measures}

The three medication scenarios were designed to represent both routine and safety-relevant situations in home use. No real medication was used. All tasks relied on placebo or empty medicine containers. The scenarios followed the task risk matrix introduced in Table~\ref{tab:task_risk_matrix}. On time medication represented a high routine, lower consequence action. Delayed or missed medication represented a high routine, higher consequence situation. Wrong medication detection represented a low routine, high consequence situation where system intervention should remain visible and contestable. Table~\ref{tab:tasks} summarises the scenarios.

\begin{table}[ht]
\centering
\footnotesize
\caption{Medication scenarios used in the user study.}
\label{tab:tasks}
\begin{tabular}{p{0.5cm}p{2.2cm}p{3.2cm}p{2.3cm}p{4cm}}
\toprule
\textbf{Task} & \textbf{Scenario} & \textbf{Participant activity} & \textbf{Task risk position} & \textbf{Study focus} \\
\midrule
1 & On time medication & Complete a routine medication event with the correct container & High routine, lower consequence & Routine recognition, logging, and system initiative \\
2 & Delayed or missed medication & Respond to the reminder so that the reminder sequence can be observed & High routine, higher consequence & Reminder response, interruption, escalation, support, and autonomy \\
3 & Wrong medication detection & Use an incorrect container and respond to the system warning & Low routine, high consequence & Safety warning, trust, intervention, confirmation, and willingness to challenge the system \\
\bottomrule
\end{tabular}
\end{table}

Across all three tasks, the main experimental difference was the allocation of control between the participant and the system. In Condition A, participants confirmed recognition results before medication events were recorded. In Condition B, routine recognition and logging were completed with less user input after the participant initiated the task. An undo option preserved a recovery point after system action. In Condition C, the system completed routine actions with minimal interruption and reserved more visible feedback for safety events.

The study combined baseline measures, condition ratings, final comparison items, and qualitative feedback. Unless otherwise specified, rating items used five-point Likert scales. For composite scores, negatively worded items were reverse-coded so that higher composite scores consistently indicated more favourable perceptions. The privacy concern item (‘I am worried the system may compromise my privacy’) was reverse-coded before calculating the Privacy \& Transparency composite.

Before interacting with the system, participants completed a baseline questionnaire. It covered demographic information, medication experience, digital confidence, privacy preferences, desire to retain decision authority, and comfort with automated systems in routine health contexts. After each automation condition, participants completed a questionnaire for that condition. It assessed confidence and comfort, perceived usefulness, perceived ease of use, usability, privacy and transparency, trust and reliance, and autonomy and control. Final comparison items captured preference rankings, perceived dignity, interruption, safety, explainability, satisfaction, and views on which actions should be automated or confirmed.

These constructs were selected because the study treated automation as more than a usability issue. It also examined trust, reliance, privacy, and user control in home medication support. Perceived usefulness and perceived ease of use were adapted from the Technology Acceptance Model~\cite{davis1989user}. Usability was measured using the System Usability Scale~\cite{brooke1996sus}. Trust and reliance items were informed by trust in automation~\cite{lee2004trust} and appropriate reliance literature. Privacy, transparency, autonomy, and control items were adapted from ethical acceptance and smart home healthcare literature. Their wording was tailored to medication recognition, logging, confirmation, undo, and data sharing. Example items and their literature basis are summarised in Table~\ref{tab:measures_constructs}. The full questionnaire is provided in Appendix~\ref{app:questionnaire}.

\begin{table}[ht!]
\centering
\footnotesize
\caption{Questionnaire constructs, example items, and literature basis.}
\label{tab:measures_constructs}
\begin{tabular}{p{3cm}p{6cm}p{4cm}}
\toprule
\textbf{Construct} & \textbf{Example items} & \textbf{Literature basis} \\
\midrule
Confidence and comfort
& I felt confident completing the medication task. I felt comfortable relying on the system.
& Health technology acceptance and comfort in use~\cite{alquran2024impact,chamorro2021self} \\

Perceived usefulness
& This system helps me manage medication effectively. It improves my medication adherence.
& Technology Acceptance Model~\cite{davis1989user} \\

Perceived ease of use
& The system is easy to learn. Interactions are clear and understandable. I find the system easy to use.
& Technology Acceptance Model~\cite{davis1989user} \\

Usability
& The ten item System Usability Scale, covering system complexity, ease of use, learnability, integration, and confidence.
& System Usability Scale~\cite{brooke1996sus} \\

Privacy and transparency
& I understand what data the system collects and why. I can control whether data is stored locally or in the cloud. I can review or delete my data easily.
& Ethical acceptance and smart home healthcare literature~\cite{felber2023mapping,pirzada2022ethics,zhu2022ethical} \\

Trust and reliance
& I trust the system to identify medications accurately. I knew when I should rely on the system and when I should not. The system encouraged an appropriate level of reliance.
& Trust in automation and appropriate reliance~\cite{lee2004trust} \\

Autonomy and control
& I can confirm decisions before they are executed. I can undo or override system actions easily. I feel in control of the medication process.
& Meaningful human control and ethical acceptance literature~\cite{santoni2018meaningful,pirzada2022ethics} \\
\bottomrule
\end{tabular}
\end{table}

\subsection{Procedure}

Each participant took part in one study session in a controlled research setting. After arrival, the researcher explained the purpose and structure of the study, answered questions, confirmed eligibility, and obtained informed consent. Participants were reminded that participation was voluntary. They could pause or withdraw at any time. They were also told that the study used simulated medication scenarios rather than real medication.

Participants first completed the baseline questionnaire. They then received a brief orientation to the Smart Medication Support system. The orientation introduced the general interface, basic task initiation, reminder delivery, recognition result display, and feedback presentation. It did not rehearse the task sequence. It also did not disclose condition order or demonstrate condition-specific behaviours in advance.

Participants completed three scripted medication scenarios within each automation condition. 
The order of automation condition blocks was counterbalanced across participants using a balanced Latin-square schedule, as evenly as recruitment allowed, to reduce learning, fatigue, and novelty effects associated with the automation manipulation. 
Within each automation block, the scenario sequence was held constant: on-time medication, delayed or missed medication, and wrong-medication detection. This fixed sequence was chosen because the scenarios differed in risk and interaction complexity, progressing from a routine lower-consequence task to a higher-consequence reminder task and then to a safety-relevant wrong-medication task. Keeping this scenario progression constant ensured that each automation condition was evaluated against the same task context, and avoided priming routine tasks with the safety-warning scenario. Accordingly, we treat automation level as the primary within-subjects manipulation, while scenario responses are interpreted as contextual task responses rather than as effects independent of task order.

After each condition, participants completed the condition questionnaire and responded to brief follow-up prompts. After completing all conditions, they completed the final comparison items and open questions about preferred automation boundaries and desired system improvements. The older adult interview subset then took part in a short semi-structured interview. The interview focused on perceived control, trust, privacy concerns, caregiver involvement, and views on which medication actions should be automated, confirmed, or reviewed. Participants were then thanked and received the approved reimbursement.

\subsection{Data Analysis}

The analysis used both quantitative and qualitative data. Quantitative data included baseline questionnaire responses, ratings after each automation condition, final comparison ratings, preference rankings, and task records. Qualitative data included open-text responses, brief comments after each condition, observation notes, and interview responses where available.

We used descriptive statistics to summarise participant characteristics, task outcomes, and questionnaire responses within each automation condition. For participants who completed all three conditions, automation level was treated as a repeated factor. For statistical comparisons, the main analyses focused on participants who completed all three automation conditions. Because the condition-level ratings were based on ordinal Likert-type items, we used Friedman tests to examine overall differences across the three automation conditions. When Friedman tests~\cite{pereira2015overview} indicated a significant overall difference, we conducted Bonferroni-corrected Wilcoxon signed-rank tests~\cite{pereira2015overview,zimmerman1993relative} for pairwise post-hoc comparisons. Kendall's $W$~\cite{abdi2007kendall} was reported as an effect size for Friedman tests. For multi-item scales, only participants with complete responses for the relevant item set were included. Interview transcripts from 11 older adult participants were analysed thematically. We identified automation preferences, perceived risks, privacy concerns, control expectations, and dignity concerns, and then grouped recurring patterns into themes. The qualitative analysis was used to explain why participants accepted or rejected different automation boundaries rather than to statistically generalise across all older adults.

\section{Results}

A total of 53 questionnaire responses were collected, and 44 were marked as complete. Among the completed responses, 32 participants completed all three automation conditions, while the remaining 12 completed only a single-condition version of the survey. Because direct comparison across automation levels requires within-participant data, the analyses reported below focus on the 32 participants who completed all three conditions. For some scales, item non-response reduced the number of complete cases to 27. Semi-structured interviews with 11 older adult participants were analysed to contextualise the questionnaire findings, particularly around control, privacy, accessibility, and acceptable automation boundaries.

The dataset consisted of questionnaire exports, supported by observation notes and open responses. The results focus on questionnaire outcomes, cross-condition comparisons, preference rankings, and qualitative observations. The full questionnaire is provided in Appendix~\ref{app:questionnaire}.

\subsection{Usability and Technology Acceptance}

Usability and technology acceptance measures showed a consistent advantage for Condition B (Auto with Undo). These measures included confidence and comfort, perceived usefulness, perceived ease of use, and the System Usability Scale (SUS). Perceived usefulness and perceived ease of use were adapted from the Technology Acceptance Model (TAM)~\cite{davis1989technology}, while usability was measured using SUS~\cite{brooke1996sus}. Across participants with complete condition-level scale data ($n = 27$), Condition B received the highest ratings for confidence and comfort ($M = 4.02, SD = 0.69$), followed by Condition A ($M = 3.78, SD = 0.76$) and Condition C ($M = 3.52, SD = 0.86$). A Friedman test indicated a significant overall difference across conditions, $\chi^2(2) = 20.02, p < .001$, with a moderate effect size, Kendall's $W = .37$. Bonferroni-corrected Wilcoxon signed-rank post-hoc tests showed that Condition B was rated significantly higher than Condition C ($p < .001$).

\begin{table*}[h!]
\centering
\small
\caption{Ethical perception results across automation conditions, covering privacy and transparency, trust and appropriate reliance, and autonomy and control (complete cases, n = 27). Values are reported as mean (SD). For the Privacy \& Transparency composite, the privacy concern item was reverse-coded so that higher scores indicate more favourable perceived privacy/transparency. Friedman test results and Kendall’s W effect sizes are shown for overall condition differences.}
\label{tab:usability_results}
\begin{tabular}{lcccc}
\toprule
\textbf{Measure} & \textbf{Condition A} & \textbf{Condition B} & \textbf{Condition C} & \textbf{Friedman test and effect size} \\
\midrule
Confidence \& Comfort & 3.78 (0.76) & 4.02 (0.69) & 3.52 (0.86) & $\chi^2(2) = 20.02, p < .001, W = .37$ \\
Perceived Usefulness & 3.72 (0.84) & 3.92 (0.87) & 3.79 (0.98) & $\chi^2(2) = 1.80, p = .406, W = .03$ \\
Perceived Ease of Use & 3.78 (0.88) & 3.96 (0.71) & 3.97 (0.86) & $\chi^2(2) = 4.79, p = .091, W = .09$ \\
SUS Score & 64.91 (17.62) & 68.06 (18.89) & 61.67 (19.84) & $\chi^2(2) = 4.64, p = .098, W = .09$ \\
\bottomrule
\end{tabular}
\end{table*}

Other usability and technology acceptance measures followed the same direction but did not reach statistical significance. Condition B had the highest mean score for perceived usefulness and SUS, while perceived ease of use was almost identical between Conditions B and C. These results suggest that participants responded most positively to the medium-automation condition in terms of confidence and comfort. Other usability and technology-acceptance measures showed the same descriptive tendency, including a lower SUS mean for Condition C, but these differences did not reach statistical significance.

\subsection{Ethical Perception: Privacy, Trust, and Autonomy}

The condition-level privacy, trust, and autonomy measures showed a similar overall pattern. These measures captured whether participants perceived the system as understandable, supportive of appropriate reliance, and compatible with user control during medication tasks. For the Privacy \& Transparency composite, with the privacy concern item reverse-coded, Condition B received the highest mean score ($M = 3.75, SD = 0.61$), followed by Condition A ($M = 3.71, SD = 0.57$) and Condition C ($M = 3.47, SD = 0.74$). A Friedman test indicated a significant overall difference across conditions, $\chi^2(2) = 7.30, p = .026$, with a small-to-moderate effect size, Kendall's $W = .14$. Bonferroni-corrected Wilcoxon signed-rank post-hoc tests showed that Condition B was rated significantly higher than Condition C ($p = .005$).

\begin{table*}[h!]
\centering
\small
\caption{Ethical perception results across automation conditions, covering privacy and transparency, trust and reliance, and autonomy and control (complete cases, $n = 27$). Values are reported as mean (SD). Friedman test results and Kendall's $W$ effect sizes are shown for overall condition differences.}
\label{tab:pta_results}
\begin{tabular}{lcccc}
\toprule
\textbf{Measure} & \textbf{Condition A} & \textbf{Condition B} & \textbf{Condition C} & \textbf{Friedman test and effect size} \\
\midrule
Privacy \& Transparency & 3.71 (0.57) & 3.75 (0.61) & 3.47 (0.74) & $\chi^2(2) = 7.30, p = .026, W = .14$ \\
Trust and Reliance & 3.86 (0.78) & 4.00 (0.67) & 3.72 (0.68) & $\chi^2(2) = 7.71, p = .021, W = .14$ \\
Autonomy and Control & 3.86 (0.97) & 4.01 (0.63) & 3.14 (1.11) & $\chi^2(2) = 9.48, p = .009, W = .18$ \\
\bottomrule
\end{tabular}
\end{table*}

Trust and reliance were also highest in Condition B, with a significant overall difference across conditions, $\chi^2(2) = 7.71, p = .021, W = .14$. Because this composite included both direct trust items and an item about double-checking system output, it should be interpreted as perceived calibrated reliance rather than as unconditional trust alone. Bonferroni-corrected Wilcoxon signed-rank post-hoc tests showed a significant difference between Conditions B and C ($p = .004$), while differences involving Condition A were not significant after correction. The clearest difference appeared in autonomy and control, where both Condition A and Condition B were rated significantly higher than Condition C ($p = .006$ and $p = .003$, respectively). These results indicate that participants did not reject automation itself. Rather, they responded most positively to automation that remained understandable, reversible, and compatible with a continued sense of control.

\subsection{Comparison Across Conditions}

The direct comparison items reinforced the same pattern. Among the 32 participants who completed all three conditions, Condition B was most often ranked as the preferred condition ($n = 17$), followed by Condition A ($n = 12$), while only three participants ranked Condition C first. Mean preference ranks were 1.53 for Condition B, 1.84 for Condition A, and 2.63 for Condition C. A Friedman test confirmed a significant difference in preference rank, $\chi^2(2) = 20.31, p < .001$, with a moderate effect size, Kendall's $W = .32$. Bonferroni-corrected Wilcoxon signed-rank post-hoc tests showed that both Condition A and Condition B were preferred significantly over Condition C ($p = .014$ and $p < .001$, respectively), while the difference between A and B was not significant after correction.

\begin{table*}[h!]
\centering
\small
\caption{Cross-condition comparison ratings among participants who completed all three conditions ($n = 32$). Values are reported as mean (SD). Friedman test results and Kendall's $W$ effect sizes are shown for overall condition differences. For the interruption/intrusiveness item, higher scores indicate greater perceived interruption.}
\label{tab:comparison_results}
\begin{tabular}{>{\raggedright\arraybackslash}p{3cm}cccc}
\toprule
\textbf{Comparison item} & \textbf{Condition A} & \textbf{Condition B} & \textbf{Condition C} & \textbf{Friedman test and effect size} \\
\midrule
Dignity and Respect & 3.97 (1.03) & 4.12 (0.87) & 3.34 (1.21) & $\chi^2(2) = 10.43, p = .005, W = .16$ \\
Safety and Reassurance & 4.22 (0.94) & 4.09 (0.89) & 3.22 (1.10) & $\chi^2(2) = 17.59, p < .001, W = .27$ \\
Transparency and Explainability & 4.25 (0.80) & 4.00 (0.95) & 3.22 (0.97) & $\chi^2(2) = 20.10, p < .001, W = .31$ \\
Overall Satisfaction & 4.06 (0.95) & 4.31 (0.74) & 3.22 (1.16) & $\chi^2(2) = 28.13, p < .001, W = .44$ \\
Interruption / Intrusiveness & 3.06 (1.05) & 3.31 (1.09) & 2.62 (1.13) & $\chi^2(2) = 11.82, p = .003, W = .18$ \\
\bottomrule
\end{tabular}
\end{table*}

Detailed pairwise post-hoc results are provided in Appendix Table~\ref{tab:posthoc_comparison}. Across dignity and respect, safety and reassurance, transparency and explainability, and overall satisfaction, Conditions A and B were consistently rated significantly higher than Condition C, while differences between A and B were not significant after correction. This pattern suggests that participants valued both explicit confirmation and auto-with-undo more than fully automatic support for ethically sensitive aspects of medication assistance.

One exception concerned interruption or intrusiveness. Condition C had the lowest interruption score ($M = 2.62, SD = 1.13$), while Condition B had the highest ($M = 3.31, SD = 1.09$). Bonferroni-corrected Wilcoxon signed-rank post-hoc tests showed that Condition B was rated significantly higher than Condition C ($p = .001$), while the differences between Conditions A and C and between Conditions A and B were not significant after correction. This lower interruption did not translate into higher preference, trust, or satisfaction. In other words, participants did not simply prefer the least disruptive system. They preferred the system that offered the most acceptable balance between convenience and control.

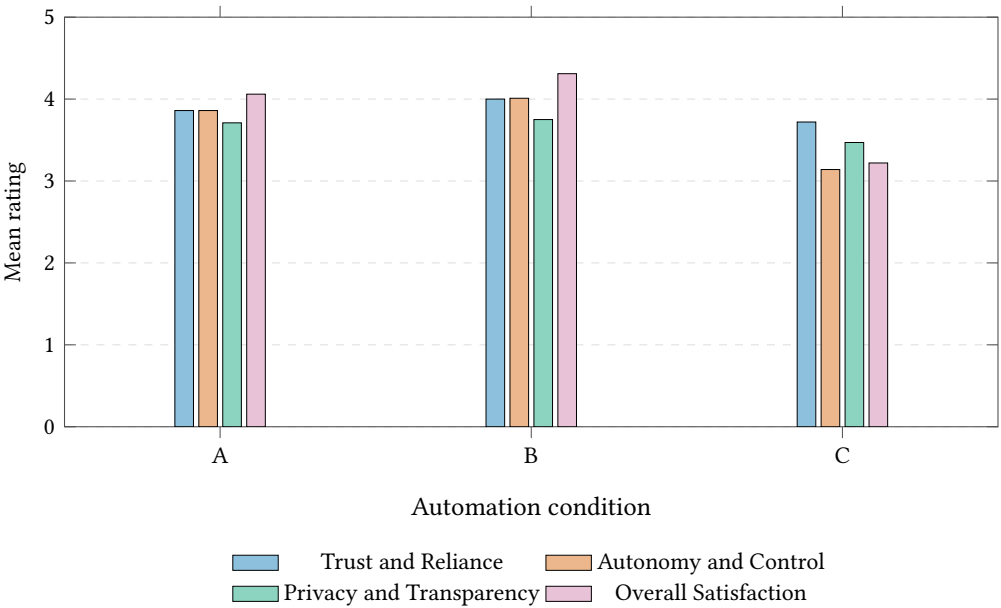
\begin{figure}[ht]
\centering
\begin{tikzpicture}
\begin{axis}[
width=\linewidth,
height=7cm,
ybar,
bar width=7pt,
ymin=0,
ymax=5,
ylabel={Mean rating},
xlabel={Automation condition},
symbolic x coords={A,B,C},
xtick=data,
xticklabels={A,B,C},
x tick label style={font=\small},
tick label style={font=\small},
xlabel style={yshift=-0.2cm},
ylabel style={font=\small},
ymajorgrids=true,
grid style={dashed, gray!25},
axis line style={black!70},
tick style={black!70},
enlarge x limits=0.25,
legend style={
at={(0.5,-0.28)},
anchor=north,
legend columns=2,
draw=none,
font=\small
},
area legend
]

\addplot+[
fill=TrustBlue!45,
draw=black,
line width=0.35pt
] coordinates {
(A,3.86)
(B,4.00)
(C,3.72)
};
\addlegendentry{Trust and Reliance}

\addplot+[
fill=AutonomyOrange!45,
draw=black,
line width=0.35pt
] coordinates {
(A,3.86)
(B,4.01)
(C,3.14)
};
\addlegendentry{Autonomy and Control}

\addplot+[
fill=PrivacyGreen!45,
draw=black,
line width=0.35pt
] coordinates {
(A,3.71)
(B,3.75)
(C,3.47)
};
\addlegendentry{Privacy and Transparency}

\addplot+[
fill=SatisfactionPurple!45,
draw=black,
line width=0.35pt
] coordinates {
(A,4.06)
(B,4.31)
(C,3.22)
};
\addlegendentry{Overall Satisfaction}

\end{axis}
\end{tikzpicture}
\caption{Mean trust, autonomy, privacy, and satisfaction ratings across the three automation conditions. A = Confirmation Required; B = Auto with Undo; C = Fully Automatic. Privacy and transparency, trust and reliance, and autonomy and control are based on complete condition-level scale cases ($n = 27$), while overall satisfaction is based on participants who completed all three conditions ($n = 32$). The grouped comparison shows that ratings did not improve monotonically from Condition A to Condition C.}
\label{fig:automation_trend}
\end{figure}

Figure~\ref{fig:automation_trend} summarizes condition-wise differences in trust and ethical acceptance across the three automation conditions. The grouped ratings show that participants did not simply respond more positively from Condition A to Condition C. Instead, trust, autonomy, privacy and transparency, and overall satisfaction were highest or near highest in Condition B, while the fully automatic condition showed a consistent decline across these dimensions.

The trust and acceptance measures varied in a similar direction. Condition B combined the highest trust and reliance score with the highest overall satisfaction and dignity ratings. Condition C was lower not only in autonomy and control, but also in trust, dignity, safety, transparency, and satisfaction. This suggests that trust was connected to whether the system remained understandable, respectful, and open to user correction.

\subsection{Qualitative Findings}

The open responses and interviews with 11 older adult participants helped explain why participants preferred different automation conditions. In particular, the qualitative data were used to interpret minority or unexpected patterns in the quantitative results, such as why some participants still preferred Condition C even though the fully automatic condition received lower overall ratings for autonomy, trust, transparency, and satisfaction. Table~\ref{tab:older_adult_interview_summary} summarises the older adult interview findings.

\begin{table*}[h!]
\centering
\tiny
\caption{Summary of semi-structured interviews with 11 older adult participants. A = Confirmation Required; B = Auto with Undo; C = Fully Automatic.}
\label{tab:older_adult_interview_summary}
\begin{tabular}{p{0.25cm}p{0.5cm}p{3.8cm}p{3.8cm}p{3.8cm}}
\toprule
\textbf{ID} & \textbf{Prefer} & \textbf{Main reason} & \textbf{Main concern} & \textbf{Design implication} \\
\midrule
P01 & C & Valued automatic recognition and recording for busy routines. & Wanted confirmation for abnormal cases, prescription changes, and data sharing. & Automate routine actions, but confirm exceptions. \\
P02 & B & Felt safest when reminders were combined with user checking. & Worried about misrecognition or incorrect intake records. & Keep intake records, wrong-medication detection, and dosage changes confirmable. \\
P03 & A & Valued visible control and local handling of medication data. & Concerned about continuous monitoring and hidden data transmission. & Provide local storage, sharing prompts, and visible privacy controls. \\
P04 & C & Reduced memory burden and allowed family to monitor medication records. & Repeated confirmations and small-screen interactions were difficult. & Reduce manual steps while strengthening audio and visual feedback. \\
P05 & B & Preferred dual verification between system assistance and user confirmation. & Feared handing medication safety decisions to code. & Automate routine recognition and reminders, but confirm safety-critical actions. \\
P06 & C & Valued seamless automation and convenience for routine medication. & Still wanted confirmation for possible wrong-medication cases. & Allow routine automation while confirming risk interventions. \\
P07 & A & Preferred a tool that acts only when needed. & Rejected automatic upload, continuous monitoring, and network transmission. & Offer offline or no-network modes for privacy-sensitive users. \\
P08 & C & Reduced button pressing, memory burden, and reliance on family monitoring. & Felt uncertain when automatic recording feedback was not repeated clearly. & Provide voice feedback, large text, and repeated confirmation of recorded actions. \\
P09 & B & Saw the system as an electronic checker rather than a decision maker. & Worried that accidental triggering could create false medication records. & Add confirmation and emergency undo for medication status records. \\
P10 & C & Valued automatic tracking and integration with health management services. & Wanted final authority over prescription dosage changes and online payments. & Separate routine automation from high-stakes medical or financial decisions. \\
P11 & A & Manual and local control because connected recording caused anxiety. & Worried about black-box uploading, data leakage, and loss of control. & Provide local-only operation, data-flow explanations, and a physical offline switch. \\
\bottomrule
\end{tabular}
\end{table*}

Three qualitative patterns help explain the quantitative results. First, Condition C was not experienced as a simple loss of control by all participants. Although it was least preferred overall, several older adult interviewees favoured the fully automatic condition because it reduced memory and physical burden. Participants who mentioned poor memory, weak vision, hand tremor, difficulty pressing small buttons, or reliance on family monitoring valued less checking, less button pressing, and fewer steps to remember. For these participants, repeated confirmation could become another task burden rather than a source of safety. This explains why fully automatic support remained acceptable for some users, even though it received lower overall ratings for autonomy, trust, and transparency.

Second, participants who preferred Condition A framed automation risk in terms of agency and privacy. They were concerned about automatic recording, continuous data capture, cloud storage, and unclear data transfer. For these participants, the key issue was not ease of use, but whether the system acted beyond their awareness or shared information without clear consent. This helps explain why full automation could be experienced as less interruptive but still less acceptable: a system can feel quiet while also feeling intrusive if users do not understand what it records, where the data goes, or who can access it.

Third, participants who preferred Condition B described undo as a way to balance support and control. The system could help with reminders, recognition, and recording, while users could still check, correct, or undo system actions. Undo therefore worked not only as an error-recovery feature, but also as a reassurance mechanism. It made automation feel less final and helped participants accept system initiative without feeling that medication decisions had been fully handed over to the machine.

The interviews suggest that acceptable automation is not only about the level of automation. It also depends on users' capabilities, privacy orientation, care context, and ability to recover from system errors. Routine actions can be automated when they are clear and reversible, while actions related to safety, prescription changes, and data sharing should require clear user confirmation.

\section{Discussion}

Our findings show that automation in home medication support is not only about convenience. Participants also judged each condition by trust, clarity, dignity, and control. The most accepted design was not the most automated one. It was the design that reduced routine effort but still allowed users to correct the system. This suggests that automation boundaries are also ethical boundaries. The key question is not how much the system can automate, but when the system should act, when it should ask, and when users should review its actions.

The results support a task-risk logic for smart medication systems. Participants accepted automation more easily for routine reminders and simple support. These actions reduced memory burden and did not take away major decisions. In contrast, wrong-medication warnings, dosage changes, data sharing, and medication records were often seen as needing confirmation, review, or undo. Low-risk routine actions can be automated with clear feedback. Routine but important actions, such as medication logging, should provide undo or review. High-risk actions, such as prescription changes or sharing records, should require explicit confirmation.

The older adult interviews show that automation preferences are personal. Condition C was least preferred overall, but some older adults still liked fully automatic support. They valued it because it reduced memory burden, physical effort, repeated checking, and reliance on family members. For users with poor memory, weak vision, hand tremor, or difficulty using small screens, higher automation can reduce interaction burden. For other users, the same automation can feel like a loss of control, especially when it involves continuous data capture, cloud storage, automatic recording, or unclear data sharing. This means automation boundaries should not be based only on average preference. Smart medication systems should support adjustable automation settings. Users who need more help may prefer automatic reminders and routine records. Users who value control may need local storage, confirmation prompts, clear data explanations, and strong undo options. Appropriate automation should fit both the task and the person.

\section{Limitations and Future Work}

This study has several limitations. First, the evaluation was conducted with a controlled prototype in a research setting rather than through long-term deployment in participants' homes. Although the scenarios were designed to reflect realistic medication situations, the findings mainly capture participants' immediate responses to the prototype. Future work should examine how trust, reliance, privacy concerns, and acceptance evolve through repeated use, changing routines, technical failures, and real medication consequences.

Second, the analysis relied mainly on questionnaires, interviews, and observation notes. These data provide useful evidence about perceived usability, trust, autonomy, privacy, comfort, and preference, but they offer limited insight into detailed interaction behaviour. Future studies should combine self-reported measures with behavioural logs to examine how perceived trust relates to actual use.

Third, the sample included a broad adult population and several stakeholder groups. This helped capture diverse perspectives on smart medication support, but it limits how specifically the findings can be generalised to older adults. The sample size was also limited, especially older adults managing multiple medications and caregivers involved in shared medication routines. Future work should also explore adaptive automation mechanisms that adjust confirmation, explanation, and undo options according to task risk and user needs.

\section{Conclusion}

This paper examines how different levels of automation shape user experience and ethical acceptance in smart medication support. Through the design and evaluation of a Smart Medication Support system, we show that automation is experienced not only through convenience and usability, but also through trust, autonomy, transparency, dignity, and comfort. Participants responded most positively to automation that reduced routine effort while preserving opportunities to review, confirm, or correct system actions. Automation boundaries should be adjusted according to task context, risk level, and user needs. Routine support may be suitable for greater automation, while actions involving medication safety, data sharing, or changes to care decisions require clearer feedback and stronger user oversight. The study contributes to HCI and smart home healthcare by framing automation as an ethical design problem rather than only a usability or efficiency issue. It also provides design implications for smart medication systems that balance convenience, safety, trust, and user control in everyday home care.

\bibliographystyle{ACM-Reference-Format}
\bibliography{reference}

\appendix
\section{Research Methods}
\subsection{Evaluation Protocol Summary}
\label{app:protocol}

This appendix provides a concise summary of the evaluation protocol used in the Smart Medication Support system study.

\textbf{Study Purpose:} The study examines how different levels of automation in a Smart Medication Support system influence user experience and ethical acceptance. The evaluation focuses on privacy, autonomy, trust, dignity, safety, comfort, and preferences regarding acceptable automation boundaries in medication-related tasks.

\textbf{Study Design:} The study uses a within-subject design in which participants interact with the Smart Medication Support system under three automation conditions:

\begin{itemize}
    \item \textbf{Condition A: Low automation / confirmation required.} Routine medication-related actions require explicit user confirmation before completion.
    \item \textbf{Condition B: Medium automation / auto with undo.} Routine actions are completed automatically, but users are given an opportunity to undo the action.
    \item \textbf{Condition C: High automation / fully automatic.} Routine actions are completed automatically with minimal interruption, while higher-risk events still trigger system intervention.
\end{itemize}

Condition order was counterbalanced across participants as evenly as recruitment allowed. The three medication scenarios were presented in a fixed scripted sequence within each automation condition: on-time medication, delayed or missed medication, and wrong-medication detection. This sequence followed the task-risk progression of the study and was kept constant so that each automation condition was evaluated against the same task context.

\textbf{Participants and Setting:} The study is conducted in a controlled research environment. Participants include adults with relevant perspectives on medication management, including older adults, caregivers, and healthcare-related participants. All participants provide informed consent before taking part.

\textbf{Materials:} The study uses the Smart Medication Support system prototype, placebo or empty medication containers, task prompts, observation sheets, and questionnaire materials. No real medication is used during the evaluation.

\textbf{Task Scenarios:} Participants complete three task scenarios under each automation condition:

\begin{enumerate}
    \item \textbf{On-time medication task.} The participant completes a routine medication interaction using the correct medicine container.
    \item \textbf{Missed or delayed medication task.} The participant delays responding so that reminder escalation and response behaviour can be observed.
    \item \textbf{Wrong medication task.} The participant is presented with an incorrect medicine container and the system issues a warning or intervention.
\end{enumerate}

These tasks are designed to represent both routine and safety-relevant medication situations.

\textbf{Session Procedure:} Each session follows the same overall structure:

\begin{enumerate}
    \item Welcome, consent, and study explanation
    \item Baseline questionnaire
    \item Brief non-task-specific orientation to the prototype
    \item Task completion under the three automation conditions
    \item Condition-level questionnaires after each condition
    \item Final cross-condition comparison questions and open-ended feedback
\end{enumerate}

Participants are informed that they can pause or withdraw at any time.

\textbf{Data Collected:} The study collects both quantitative and qualitative data. Quantitative data include questionnaire responses, condition ratings, preference rankings, and scale scores related to usability, technology acceptance, privacy, trust, and autonomy. Qualitative data include open-ended questionnaire responses and researcher observation notes.

\textbf{Analysis Overview:} The evaluation uses a mixed-methods approach. Quantitative analyses compare condition-level ratings across the three automation conditions, while qualitative responses are analysed thematically to identify recurring concerns and preferences related to trust, control, privacy, transparency, safety, and acceptable automation boundaries.

\subsection{Questionnaire Items and Scales}
\label{app:questionnaire}

Unless otherwise noted, all Likert-scale items used a 5-point response format:
1 = Strongly disagree, 2 = Disagree, 3 = Neither agree nor disagree, 4 = Agree, and 5 = Strongly agree.

\begin{longtable}{p{0.25\linewidth} p{0.05\linewidth} p{0.65\linewidth}}
\caption{Complete questionnaire items and scales used in the deployed survey.}
\label{tab:questionnaire_all}\\
\toprule
\textbf{Section} & \textbf{No.} & \textbf{Item} \\
\midrule
\endfirsthead

\toprule
\textbf{Section} & \textbf{No.} & \textbf{Item} \\
\midrule
\endhead

\multicolumn{3}{l}{\textbf{Screening / session setup}} \\[2pt]
Setup & 0.1 & How many system versions would you like to evaluate? (e.g., complete all / single-condition version) \\

\addlinespace
\multicolumn{3}{l}{\textbf{Baseline questionnaire}} \\[2pt]
Baseline & 1.1 & Please select your age group. \\
Baseline & 1.2 & Please select your gender. \\
Baseline & 1.3 & Please select the option that best describes you. \\
Baseline & 1.3a & If other, please specify. \\
Baseline & 1.4 & How many medications do you currently take regularly? \\
Baseline & 1.5.1 & I feel confident using smart devices to manage my health. \\
Baseline & 1.5.2 & I value privacy more than convenience. \\
Baseline & 1.5.3 & I prefer to retain the final decision-making authority in critical health decisions. \\
Baseline & 1.5.4 & I have previously used medication reminder devices or apps. \\
Baseline & 1.5.5 & I am willing to share some data in exchange for improved safety and convenience. \\
Baseline & 1.6 & Please briefly describe any experience you have with medication management, for yourself or for another person. \\

\addlinespace
\multicolumn{3}{l}{\textbf{Condition A: Confirmation Required}} \\[2pt]
A: Confidence \& Comfort & 2.1.1 & I felt confident completing the medication task. \\
A: Confidence \& Comfort & 2.1.2 & I felt comfortable relying on the system. \\
A: Confidence \& Comfort & 2.1.3 & The system's behaviour matched my expectations. \\

A: Perceived Usefulness & 2.2.1 & This system helps me manage medication effectively. \\
A: Perceived Usefulness & 2.2.2 & It improves my medication adherence. \\
A: Perceived Usefulness & 2.2.3 & It enhances my overall health management. \\
A: Perceived Usefulness & 2.2.4 & I would use this system regularly. \\

A: Perceived Ease of Use & 2.3.1 & The system is easy to learn. \\
A: Perceived Ease of Use & 2.3.2 & Interactions are clear and understandable. \\
A: Perceived Ease of Use & 2.3.3 & I can quickly become skillful at using it. \\
A: Perceived Ease of Use & 2.3.4 & I find the system easy to use. \\

A: SUS & 2.4.1 & I think that I would like to use this system frequently. \\
A: SUS & 2.4.2 & I found the system unnecessarily complex. \\
A: SUS & 2.4.3 & I thought the system was easy to use. \\
A: SUS & 2.4.4 & I think that I would need the support of a technical person to use this system. \\
A: SUS & 2.4.5 & I found the various functions were well integrated. \\
A: SUS & 2.4.6 & I thought there was too much inconsistency in this system. \\
A: SUS & 2.4.7 & I would imagine most people would learn to use this system very quickly. \\
A: SUS & 2.4.8 & I found the system very cumbersome to use. \\
A: SUS & 2.4.9 & I felt very confident using the system. \\
A: SUS & 2.4.10 & I needed to learn a lot of things before I could get going. \\

A: Privacy \& Transparency & 2.5.1 & I understand what data the system collects and why. \\
A: Privacy \& Transparency & 2.5.2 & I can control whether data is stored locally or in the cloud. \\
A: Privacy \& Transparency & 2.5.3 & I am worried the system may compromise my privacy. \\
A: Privacy \& Transparency & 2.5.4 & I can review or delete my data easily. \\
A: Privacy \& Transparency & 2.5.5 & I understood why the system gave its recommendation. \\
A: Privacy \& Transparency & 2.5.6 & The system provided enough information for me to judge whether it was correct. \\
A: Privacy \& Transparency & 2.5.7 & The system's feedback was clear and easy to interpret. \\
A: Privacy \& Transparency & 2.5.8 & When the system was wrong, it was easy to understand what went wrong. \\

A: Trust \& Reliance & 2.6.1 & I trust the system to identify medications accurately. \\
A: Trust \& Reliance & 2.6.2 & I trust the system to warn me if something is unsafe. \\
A: Trust \& Reliance & 2.6.3 & I believe the system will not misuse my data. \\
A: Trust \& Reliance & 2.6.4 & I believe the system acts in my best interest. \\
A: Trust \& Reliance & 2.6.5 & I would double-check the system's output before acting on it. \\
A: Trust \& Reliance & 2.6.6 & I knew when I should rely on the system and when I should not. \\
A: Trust \& Reliance & 2.6.7 & The system encouraged an appropriate level of reliance (neither too much nor too little). \\

A: Autonomy & 2.7.1 & I can choose which functions are automated. \\
A: Autonomy & 2.7.2 & I can confirm decisions before they are executed. \\
A: Autonomy & 2.7.3 & I can undo or override system actions easily. \\
A: Autonomy & 2.7.4 & I feel in control of the medication process. \\

A: Open response & 2.8 & Did anything about this level of automation feel too intrusive or not supportive enough? Please explain briefly. \\

\addlinespace
\multicolumn{3}{l}{\textbf{Condition B: Auto with Undo}} \\[2pt]
B: Confidence \& Comfort & 2.1.1 & I felt confident completing the medication task. \\
B: Confidence \& Comfort & 2.1.2 & I felt comfortable relying on the system. \\
B: Confidence \& Comfort & 2.1.3 & The system's behaviour matched my expectations. \\

B: Perceived Usefulness & 2.2.1 & This system helps me manage medication effectively. \\
B: Perceived Usefulness & 2.2.2 & It improves my medication adherence. \\
B: Perceived Usefulness & 2.2.3 & It enhances my overall health management. \\
B: Perceived Usefulness & 2.2.4 & I would use this system regularly. \\

B: Perceived Ease of Use & 2.3.1 & The system is easy to learn. \\
B: Perceived Ease of Use & 2.3.2 & Interactions are clear and understandable. \\
B: Perceived Ease of Use & 2.3.3 & I can quickly become skillful at using it. \\
B: Perceived Ease of Use & 2.3.4 & I find the system easy to use. \\

B: SUS & 2.4.1 & I think that I would like to use this system frequently. \\
B: SUS & 2.4.2 & I found the system unnecessarily complex. \\
B: SUS & 2.4.3 & I thought the system was easy to use. \\
B: SUS & 2.4.4 & I think that I would need the support of a technical person to use this system. \\
B: SUS & 2.4.5 & I found the various functions were well integrated. \\
B: SUS & 2.4.6 & I thought there was too much inconsistency in this system. \\
B: SUS & 2.4.7 & I would imagine most people would learn to use this system very quickly. \\
B: SUS & 2.4.8 & I found the system very cumbersome to use. \\
B: SUS & 2.4.9 & I felt very confident using the system. \\
B: SUS & 2.4.10 & I needed to learn a lot of things before I could get going. \\

B: Privacy \& Transparency & 2.5.1 & I understand what data the system collects and why. \\
B: Privacy \& Transparency & 2.5.2 & I can control whether data is stored locally or in the cloud. \\
B: Privacy \& Transparency & 2.5.3 & I am worried the system may compromise my privacy. \\
B: Privacy \& Transparency & 2.5.4 & I can review or delete my data easily. \\
B: Privacy \& Transparency & 2.5.5 & I understood why the system gave its recommendation. \\
B: Privacy \& Transparency & 2.5.6 & The system provided enough information for me to judge whether it was correct. \\
B: Privacy \& Transparency & 2.5.7 & The system's feedback was clear and easy to interpret. \\
B: Privacy \& Transparency & 2.5.8 & When the system was wrong, it was easy to understand what went wrong. \\

B: Trust \& Reliance & 2.6.1 & I trust the system to identify medications accurately. \\
B: Trust \& Reliance & 2.6.2 & I trust the system to warn me if something is unsafe. \\
B: Trust \& Reliance & 2.6.3 & I believe the system will not misuse my data. \\
B: Trust \& Reliance & 2.6.4 & I believe the system acts in my best interest. \\
B: Trust \& Reliance & 2.6.5 & I would double-check the system's output before acting on it. \\
B: Trust \& Reliance & 2.6.6 & I knew when I should rely on the system and when I should not. \\
B: Trust \& Reliance & 2.6.7 & The system encouraged an appropriate level of reliance (neither too much nor too little). \\

B: Autonomy & 2.7.1 & I can choose which functions are automated. \\
B: Autonomy & 2.7.2 & I can confirm decisions before they are executed. \\
B: Autonomy & 2.7.3 & I can undo or override system actions easily. \\
B: Autonomy & 2.7.4 & I feel in control of the medication process. \\

B: Open response & 2.8 & Did anything about this level of automation feel too intrusive or not supportive enough? Please explain briefly. \\

\addlinespace
\multicolumn{3}{l}{\textbf{Condition C: Fully Automatic}} \\[2pt]
C: Confidence \& Comfort & 2.1.1 & I felt confident completing the medication task. \\
C: Confidence \& Comfort & 2.1.2 & I felt comfortable relying on the system. \\
C: Confidence \& Comfort & 2.1.3 & The system's behaviour matched my expectations. \\

C: Perceived Usefulness & 2.2.1 & This system helps me manage medication effectively. \\
C: Perceived Usefulness & 2.2.2 & It improves my medication adherence. \\
C: Perceived Usefulness & 2.2.3 & It enhances my overall health management. \\
C: Perceived Usefulness & 2.2.4 & I would use this system regularly. \\

C: Perceived Ease of Use & 2.3.1 & The system is easy to learn. \\
C: Perceived Ease of Use & 2.3.2 & Interactions are clear and understandable. \\
C: Perceived Ease of Use & 2.3.3 & I can quickly become skillful at using it. \\
C: Perceived Ease of Use & 2.3.4 & I find the system easy to use. \\

C: SUS & 2.4.1 & I think that I would like to use this system frequently. \\
C: SUS & 2.4.2 & I found the system unnecessarily complex. \\
C: SUS & 2.4.3 & I thought the system was easy to use. \\
C: SUS & 2.4.4 & I think that I would need the support of a technical person to use this system. \\
C: SUS & 2.4.5 & I found the various functions were well integrated. \\
C: SUS & 2.4.6 & I thought there was too much inconsistency in this system. \\
C: SUS & 2.4.7 & I would imagine most people would learn to use this system very quickly. \\
C: SUS & 2.4.8 & I found the system very cumbersome to use. \\
C: SUS & 2.4.9 & I felt very confident using the system. \\
C: SUS & 2.4.10 & I needed to learn a lot of things before I could get going. \\

C: Privacy \& Transparency & 2.5.1 & I understand what data the system collects and why. \\
C: Privacy \& Transparency & 2.5.2 & I can control whether data is stored locally or in the cloud. \\
C: Privacy \& Transparency & 2.5.3 & I am worried the system may compromise my privacy. \\
C: Privacy \& Transparency & 2.5.4 & I can review or delete my data easily. \\
C: Privacy \& Transparency & 2.5.5 & I understood why the system gave its recommendation. \\
C: Privacy \& Transparency & 2.5.6 & The system provided enough information for me to judge whether it was correct. \\
C: Privacy \& Transparency & 2.5.7 & The system's feedback was clear and easy to interpret. \\
C: Privacy \& Transparency & 2.5.8 & When the system was wrong, it was easy to understand what went wrong. \\

C: Trust \& Reliance & 2.6.1 & I trust the system to identify medications accurately. \\
C: Trust \& Reliance & 2.6.2 & I trust the system to warn me if something is unsafe. \\
C: Trust \& Reliance & 2.6.3 & I believe the system will not misuse my data. \\
C: Trust \& Reliance & 2.6.4 & I believe the system acts in my best interest. \\
C: Trust \& Reliance & 2.6.5 & I would double-check the system's output before acting on it. \\
C: Trust \& Reliance & 2.6.6 & I knew when I should rely on the system and when I should not. \\
C: Trust \& Reliance & 2.6.7 & The system encouraged an appropriate level of reliance (neither too much nor too little). \\

C: Autonomy & 2.7.1 & I can choose which functions are automated. \\
C: Autonomy & 2.7.2 & I can confirm decisions before they are executed. \\
C: Autonomy & 2.7.3 & I can undo or override system actions easily. \\
C: Autonomy & 2.7.4 & I feel in control of the medication process. \\

C: Open response & 2.8 & Did anything about this level of automation feel too intrusive or not supportive enough? Please explain briefly. \\

\addlinespace
\multicolumn{3}{l}{\textbf{Cross-condition comparison}} \\[2pt]
Comparison & 5.1.1 & Preferred Condition ranking: Condition A -- Confirmation Required \\
Comparison & 5.1.2 & Preferred Condition ranking: Condition B -- Auto with Undo \\
Comparison & 5.1.3 & Preferred Condition ranking: Condition C -- Fully Automatic \\

Comparison & 5.2.1 & Condition A: Sense of respect and dignity \\
Comparison & 5.2.2 & Condition A: Level of interruption or intrusiveness \\
Comparison & 5.2.3 & Condition A: Sense of safety and reassurance \\
Comparison & 5.2.4 & Condition A: Transparency and explainability \\
Comparison & 5.2.5 & Condition A: Overall satisfaction \\

Comparison & 5.3.1 & Condition B: Sense of respect and dignity \\
Comparison & 5.3.2 & Condition B: Level of interruption or intrusiveness \\
Comparison & 5.3.3 & Condition B: Sense of safety and reassurance \\
Comparison & 5.3.4 & Condition B: Transparency and explainability \\
Comparison & 5.3.5 & Condition B: Overall satisfaction \\

Comparison & 5.4.1 & Condition C: Sense of respect and dignity \\
Comparison & 5.4.2 & Condition C: Level of interruption or intrusiveness \\
Comparison & 5.4.3 & Condition C: Sense of safety and reassurance \\
Comparison & 5.4.4 & Condition C: Transparency and explainability \\
Comparison & 5.4.5 & Condition C: Overall satisfaction \\

\addlinespace
\multicolumn{3}{l}{\textbf{Automation boundary and final feedback}} \\[2pt]
Final & 6.1 & Fully Automated Actions \\
Final & 6.2 & Confirmation Required Actions \\
Final & 6.3 & Do you have any suggestions to improve the system's trustworthiness, privacy, autonomy, or dignity? \\

\bottomrule
\end{longtable}

\subsection{Questionnaire data}

\begin{table}[h!]
\centering
\caption{Overview of the questionnaire dataset.}
\label{tab:sample_overview}
\begin{tabular}{lr}
\toprule
\textbf{Dataset} & \textbf{Value} \\
\midrule
Total responses collected & 53 \\
Completed responses & 44 \\
Incomplete responses & 9 \\
Completed all three conditions & 32 \\
Completed Condition A only & 6 \\
Completed Condition B only & 5 \\
Completed Condition C only & 1 \\
Complete cases for condition-level scale analysis & 27 \\
\bottomrule
\end{tabular}
\end{table}

\begin{table}[h!]
\centering
\caption{Participant preferences for automation boundaries in medication-related actions.}
\label{tab:automation_boundary}
\begin{tabular}{lp{5.3cm}}
\toprule
\textbf{Preference} & \textbf{Most frequently selected actions} \\
\midrule
Suitable for full automation & Sending reminders ($n=25$); routine logging ($n=18$) \\
Should require confirmation & Incorrect dosage warning ($n=20$); routine medication logging ($n=18$); wrong medication detection ($n=18$) \\
\bottomrule
\end{tabular}
\end{table}

\subsection{Detailed Condition-Level Results}
\label{app:detailed_results}

The following tables report item-level descriptive statistics for participants who completed all three automation conditions ($n=32$). Values are reported as mean (SD). Due to item-level nonresponse, the valid number of responses varies slightly across items.

\begin{table*}[h!]
\centering
\caption{Detailed condition-level usability and acceptance items. Values are reported as mean (SD).}
\label{tab:detailed_usability_items}
\begin{tabular}{p{0.5\linewidth}ccc}
\toprule
\textbf{Item} & \textbf{Condition A} & \textbf{Condition B} & \textbf{Condition C} \\
\midrule
\multicolumn{4}{l}{\textit{Confidence \& Comfort}} \\
I felt confident completing the medication task. & 4.07 (0.86) & 4.21 (0.73) & 3.67 (0.96) \\
I felt comfortable relying on the system. & 3.54 (0.92) & 4.07 (0.80) & 3.33 (1.04) \\
The system's behaviour matched my expectations. & 3.71 (0.81) & 3.93 (0.84) & 3.56 (0.93) \\

\addlinespace
\multicolumn{4}{l}{\textit{Perceived Usefulness}} \\
This system helps me manage medication effectively. & 3.86 (1.01) & 4.10 (0.94) & 3.93 (1.04) \\
It improves my medication adherence. & 3.93 (0.90) & 4.03 (0.91) & 3.93 (1.04) \\
It enhances my overall health management. & 3.68 (0.90) & 4.00 (0.85) & 3.85 (0.95) \\
I would use this system regularly. & 3.43 (1.10) & 3.69 (1.04) & 3.44 (1.19) \\

\addlinespace
\multicolumn{4}{l}{\textit{Perceived Ease of Use}} \\
The system is easy to learn. & 3.93 (1.02) & 3.93 (0.84) & 3.93 (1.04) \\
Interactions are clear and understandable. & 3.61 (0.88) & 3.90 (0.72) & 3.96 (0.85) \\
I can quickly become skillful at using it. & 3.79 (0.99) & 3.93 (0.84) & 3.89 (0.93) \\
I find the system easy to use. & 3.82 (1.06) & 4.07 (0.80) & 4.11 (0.93) \\

\addlinespace
\multicolumn{4}{l}{\textit{System Usability Scale (SUS)}} \\
I think that I would like to use this system frequently. & 3.50 (1.00) & 3.79 (1.01) & 3.56 (1.22) \\
I found the system unnecessarily complex. & 2.61 (1.07) & 2.45 (1.12) & 2.89 (1.12) \\
I thought the system was easy to use. & 4.00 (0.77) & 4.00 (0.93) & 4.11 (0.97) \\
I think that I would need the support of a technical person to use this system. & 2.54 (1.04) & 2.52 (1.18) & 3.00 (1.18) \\
I found the various functions were well integrated. & 3.96 (0.84) & 4.00 (0.93) & 3.89 (0.97) \\
I thought there was too much inconsistency in this system. & 2.43 (0.84) & 2.48 (0.95) & 2.85 (1.17) \\
I would imagine most people would learn to use this system very quickly. & 3.75 (0.97) & 4.00 (1.00) & 3.89 (1.19) \\
I found the system very cumbersome to use. & 2.86 (1.01) & 2.55 (0.99) & 2.70 (1.07) \\
I felt very confident using the system. & 3.96 (0.88) & 3.97 (0.91) & 3.74 (1.16) \\
I needed to learn a lot of things before I could get going. & 2.79 (1.07) & 2.76 (1.12) & 3.07 (1.21) \\
\bottomrule
\end{tabular}
\end{table*}

\begin{table*}[htbp]
\centering
\caption{Detailed condition-level ethical perception items. Values are reported as mean (SD).}
\label{tab:detailed_pta_items}
\begin{tabular}{p{0.5\linewidth}ccc}
\toprule
\textbf{Item} & \textbf{Condition A} & \textbf{Condition B} & \textbf{Condition C} \\
\midrule
\multicolumn{4}{l}{\textit{Privacy \& Transparency}} \\
I understand what data the system collects and why. & 3.86 (0.97) & 4.00 (0.89) & 3.56 (1.01) \\
I can control whether data is stored locally or in the cloud. & 3.61 (0.96) & 3.55 (1.09) & 3.04 (1.02) \\
I am worried the system may compromise my privacy. & 3.25 (0.97) & 3.14 (1.09) & 3.37 (1.01) \\
I can review or delete my data easily. & 3.64 (0.95) & 3.72 (1.07) & 3.15 (1.06) \\
I understood why the system gave its recommendation. & 3.89 (1.03) & 4.00 (0.93) & 3.63 (0.97) \\
The system provided enough information for me to judge whether it was correct. & 3.86 (0.65) & 4.03 (0.78) & 3.56 (1.15) \\
The system's feedback was clear and easy to interpret. & 3.89 (0.83) & 3.86 (0.95) & 4.00 (1.00) \\
When the system was wrong, it was easy to understand what went wrong. & 3.79 (0.83) & 3.90 (0.72) & 3.48 (0.98) \\

\addlinespace
\multicolumn{4}{l}{\textit{Trust \& Reliance}} \\
I trust the system to identify medications accurately. & 3.86 (0.97) & 4.00 (1.04) & 3.52 (1.31) \\
I trust the system to warn me if something is unsafe. & 3.96 (0.88) & 4.14 (0.99) & 3.85 (1.10) \\
I believe the system will not misuse my data. & 3.68 (0.86) & 3.79 (0.90) & 3.56 (0.85) \\
I believe the system acts in my best interest. & 3.82 (0.86) & 3.90 (0.90) & 3.67 (1.11) \\
I would double-check the system's output before acting on it. & 4.36 (0.91) & 4.10 (0.67) & 4.15 (0.99) \\
I knew when I should rely on the system and when I should not. & 3.68 (0.98) & 4.14 (0.74) & 3.85 (0.72) \\
The system encouraged an appropriate level of reliance (neither too much nor too little). & 3.71 (1.01) & 4.00 (1.00) & 3.48 (1.05) \\

\addlinespace
\multicolumn{4}{l}{\textit{Autonomy \& Control}} \\
I can choose which functions are automated. & 3.79 (1.10) & 3.86 (0.83) & 3.11 (1.15) \\
I can confirm decisions before they are executed. & 3.96 (1.07) & 4.21 (0.68) & 3.15 (1.20) \\
I can undo or override system actions easily. & 3.71 (1.18) & 3.86 (0.95) & 3.04 (1.22) \\
I feel in control of the medication process. & 4.00 (0.82) & 4.07 (0.84) & 3.26 (1.20) \\
\bottomrule
\end{tabular}
\end{table*}

\begin{table}[htbp]
\centering
\caption{Detailed cross-condition comparison ratings among participants who completed all three conditions ($n = 32$). Values are reported as mean (SD).}
\label{tab:detailed_comparison}
\begin{tabular}{p{0.42\linewidth}ccc}
\toprule
\textbf{Dimension} & \textbf{Condition A} & \textbf{Condition B} & \textbf{Condition C} \\
\midrule
Sense of respect and dignity & 3.97 (1.03) & 4.12 (0.87) & 3.34 (1.21) \\
Level of interruption or intrusiveness & 3.06 (1.05) & 3.31 (1.09) & 2.62 (1.13) \\
Sense of safety and reassurance & 4.22 (0.94) & 4.09 (0.89) & 3.22 (1.10) \\
Transparency and explainability & 4.25 (0.80) & 4.00 (0.95) & 3.22 (0.97) \\
Overall satisfaction & 4.06 (0.95) & 4.31 (0.74) & 3.22 (1.16) \\
\bottomrule
\end{tabular}
\end{table}

\begin{table}[htbp!]
\centering
\caption{Bonferroni-corrected Wilcoxon signed-rank post-hoc comparisons for cross-condition ratings among participants who completed all three conditions ($n = 32$).}
\label{tab:posthoc_comparison}
\begin{tabular}{lcccc}
\toprule
\textbf{Measure} & \textbf{A vs C} & \textbf{B vs C} & \textbf{A vs B} & \textbf{Interpretation} \\
\midrule
Preference Rank & $p = .014$ & $p < .001$ & n.s. & A and B were preferred over C \\
Dignity and Respect & $p = .027$ & $p = .007$ & n.s. & A and B were higher than C \\
Safety and Reassurance & $p = .002$ & $p < .001$ & n.s. & A and B were higher than C \\
Transparency and Explainability & $p < .001$ & $p = .002$ & n.s. & A and B were higher than C \\
Overall Satisfaction & $p = .003$ & $p < .001$ & n.s. & A and B were higher than C \\
Interruption / Intrusiveness & n.s. & $p = .001$ & n.s. & B was higher than C \\
\bottomrule
\end{tabular}
\begin{flushleft}
\footnotesize
Note. A = Confirmation Required; B = Auto with Undo; C = Fully Automatic. Values are Bonferroni-corrected $p$ values. n.s. = not significant after correction. For preference rank, lower rank values indicate stronger preference. For interruption/intrusiveness, higher scores indicate greater perceived interruption.
\end{flushleft}
\end{table}

\subsection{Analysis Notes}
\label{app:analysis_notes}

The dataset consists of 53 questionnaire responses, of which 44 are marked as complete. Because the study includes both full three-condition participation and single-condition completion options, different subsets of responses are used for different analyses.

For direct comparisons across automation conditions, only participants who complete all three conditions are included ($n=32$). This subset is used for preference ranking and cross-condition comparison analyses. For multi-item composite scales, such as confidence and comfort, perceived usefulness, perceived ease of use, SUS, and privacy--trust--autonomy measures, analyses are restricted to complete cases for the relevant item set. As a result, the valid sample size for some scale-level analyses is reduced to $n=27$.

Item-level descriptive statistics reported in the appendix are calculated using available responses among participants who complete all three conditions. Accordingly, the valid number of responses may vary slightly across items because of item-level nonresponse. Means and standard deviations are reported based on the available responses for each item.

Incomplete questionnaire responses are excluded from the main condition-comparison analyses. However, they are retained for descriptive review during data screening. Open-ended responses are analyzed only when participants provide substantive text. Illustrative quotations are anonymized and lightly edited for spelling and grammar where needed, without changing meaning.

\subsection{Anonymized Open-Ended Response}

Table~\ref{tab:quote_examples} is selected to illustrate recurrent themes identified across the questionnaire dataset, including preference for balanced automation, desire for continued user control in safety-relevant situations, and requests for clearer privacy communication and onboarding support.

\label{app:quotes}

\begin{table*}[ht!]
\centering
\caption{Illustrative anonymized open-ended responses.}
\label{tab:quote_examples}
\begin{tabular}{p{0.24\linewidth}p{0.12\linewidth}p{0.56\linewidth}}
\toprule
\textbf{Theme} & \textbf{Participant} & \textbf{Response} \\
\midrule
Preference for balanced automation & P01 & ``I think this is the best one to use.'' \\

Preference for balanced automation & P02 & ``Undo option made me comfortable using automation.'' \\

Preference for balanced automation & P03 & ``The confirmations improved trust without slowing me much.'' \\

Preference for balanced automation & P04 & ``Felt supportive and safe for medication use.'' \\

Value of confirmation and control & P05 & ``I liked having confirmation before actions were completed. It made me feel safer and more involved.'' \\

Value of confirmation and control & P06 & ``It was supportive and gave me control.'' \\

Concern about excessive automation & P07 & ``It was fast, but I wanted more control.'' \\

Concern about excessive automation & P08 & ``It felt a little too automatic for medication tasks.'' \\

Concern about excessive automation & P09 & ``Useful and quick, but I would still check important actions.'' \\

Concern about excessive automation & P10 & ``Maybe too intrusive because it is fully automatic, meaning no double-checking.'' \\

Need for transparency and privacy communication & P11 & ``I would like more information on how the data are used.'' \\

Need for transparency and privacy communication & P12 & ``Make the use of private information clear to the user and let them be able to access it again if needed.'' \\

Need for onboarding and clearer instructions & P13 & ``Maybe a demonstration video or manual about all the functions.'' \\

Need for accessibility improvements & P14 & ``Provide simpler explanations and larger text for older adults.'' \\

Design suggestion about automation boundaries & P15 & ``Keep safety-related actions confirmed by the user while automating simple daily tasks.'' \\
\bottomrule
\end{tabular}
\end{table*}

\end{document}